\documentclass[acmlarge]{acmart}
\makeatletter
\newcommand{\confshort}{\acmConference@shortname}
\newcommand{\conffull}{\acmConference@name}
\newcommand{\confdate}{\acmConference@date}
\newcommand{\confloc}{\acmConference@venue}
\AtBeginDocument{
  \fancypagestyle{firstpagestyle}{
    \fancyhead{}%
    \fancyfoot[C]{}%
  }
  \fancyhf{}
  \fancyhead[LO]{\@headfootfont\shorttitle}%
  \fancyhead[RE]{\@headfootfont\@shortauthors}%
  \fancyhead[LE]{\@headfootfont\footnotesize \confshort, \confdate, \confloc}%
  \fancyhead[RO]{\@headfootfont\footnotesize \confshort, \confdate, \confloc}%
  \fancyfoot[C]{}%
}
\makeatother
\acmBooktitle{\conffull\@ (\confshort), \confdate, \confloc}
\usepackage{enumitem}
\usepackage{multirow}
\usepackage{soul}
\usepackage{makecell}
\AtBeginDocument{\settopmatter{printacmref=true}}
\setcopyright{acmlicensed}
\copyrightyear{2026}
\acmYear{2026}
\setcopyright{cc}
\setcctype{by-nc-nd}
\acmConference[FAccT '26]{The 2026 ACM Conference on Fairness, Accountability, and Transparency}{June 25--28, 2026}{Montreal, QC, Canada}
\acmBooktitle{The 2026 ACM Conference on Fairness, Accountability, and Transparency (FAccT '26), June 25--28, 2026, Montreal, QC, Canada}
\acmDOI{10.1145/3805689.3806765}
\acmISBN{979-8-4007-2596-8/2026/06}
\begin{document}
\title{Challenges to Grassroots Organization Engagement with AI Policy}

\author{Carter Buckner}
\affiliation{\institution{Queer in AI}\country{USA}}
\email{carter.buckner@ostem.org}

\author{Jennifer Mickel}
\affiliation{\institution{Queer in AI}\country{USA}}
\email{jamickel@utexas.edu}

\author{Nandhini Swaminathan}
\affiliation{\institution{Queer in AI}\country{USA}}
\email{Nswaminathan@ucsd.edu}

\author{William Agnew}
\affiliation{\institution{Queer in AI}\country{USA}}
\email{wagnew@andrew.cmu.edu}

\author{Jacob Hobbs}
\affiliation{\institution{Queer in AI}\country{USA}}
\email{jacobhobbs1100@gmail.com}

\author{Sarthak Arora}
\affiliation{\institution{Queer in AI}\country{USA}}
\email{sarthakvarora@gmail.com}

\author{Michelle Lin}
\affiliation{\institution{Queer in AI}\country{CAN}}
\email{Michelle.lin2@mail.mcgill.ca}

\author{Yanan Long}
\affiliation{\institution{Queer in AI \& StickFlux Labs \& University of Chicago}\country{USA}}
\authornote{Work began while at the University of Chicago}
\email{ylong@uchicago.edu}

\author{B.V. Alaka}
\affiliation{\institution{Queer in AI}\country{USA}}
\authornote{A published version to appear at ACM FAccT 2026 mistakenly omits author}
\email{alakabv@gmail.com}

\renewcommand{\shortauthors}{C. Buckner et al.}

\begin{abstract}
Public policies are being developed around the world to address privacy, economic, intellectual property, energy, and other risks that
AI technologies pose. 
%In the US, there exists a mix of local and federal AI governance efforts with varying efforts toward public participation. 
Involvement from the general public is essential to governance as an accountability and alignment mechanism. However, participating in and impacting policymaking can be challenging for sections of the public that lack extensive networks, lobbying capabilities, and other forms of power. This challenge is especially acute for marginalized communities.
%Within traditional mechanisms aiding US federal policy advocacy, there exist many well-resourced actors
%and opaque development processes that sideline the needs of marginalized populations --- groups often lacking the extensive networks, lobbying capabilities, and other forms of power common to federal policy making \cite{rahman2017policymaking}. 
In this paper, we present a case study of our organization's efforts to bring participatory design (PD) principles to AI policymaking in the US. We describe our engagements with several US policy bodies, and our participatory development  of AI policy for queer people.
%We describe our organizational approach to large-scale PD for policy co-design and explore how PD practices fit within traditional governance mechanisms through our involvement with the National Institute for Standards and Technology (NIST), a non-regulatory US federal agency. 
%We highlight challenges of bridging PD theory and practice, surface tensions in PD for governance, and offer suggestions to alleviatethem. 
We highlight challenges with PD practice with marginalized communities, and offer suggestions to alleviate them.
We conclude with actionable recommendations for policymakers and other organizers working in marginalized communities.
\end{abstract}


\maketitle

\section{Introduction}

AI systems cause harm through bias, hallucination, misrepresentation, privacy infringement, and other mechanisms \cite{barocas2017problem,Wang2024LargeLM}. Harms from AI systems are often difficult to anticipate as not all use cases are known at deployment time \cite{Dorn2024HarmfulSD}. Marginalized groups, including the queer community, can be at greater risk and vulnerable to more types of harm \cite{chen2024lost,gillespie2024generative}. Additionally, within AI development, these same groups are often excluded, mirroring historical harms \cite{Sassaro2024TheDC,dev2021harms,keyes2018misgendering}. In response to this, there has been a surge of recent efforts to regulate AI \cite{biden_13960, trump_ai_action_2025, eu_ai_act} but, marginalized people similarly have not had their needs met by policy and have suffered misrepresentation and discriminatory harm from public governance \cite{plaxico1999,ACLU2022TSA,Romero2020LGBTPA,queerinai2023queer, Spade2011NormalLA}. 

With the recent ``participatory turn in AI design''  \cite{delgado2023participatory}, calls for empowering data and AI subjects, particularly from marginalized communities, have grown as a means of, in part, addressing issues of bias, stereotyping, and other AI harms to better align AI to the needs of users \cite{birhane2022power, corbett2023power}. However, broadening participation in AI development risks being shallow and extractive if the contributions of participants are used to improve AI systems \textit{without} the participants and their communities receiving or experiencing these benefits and having AI systems positively impact their lives \cite{maas2024beyond, birhane2022power}. Additionally, participatory design (PD) offers a pathway for equalizing public participation in governance, where calls for greater public participation are not new \cite{10.5555/3208509,gilman2022beyond,kaminski_stakeholder_2022,crump_surveillance_2016}. Importantly, mechanisms for public participation already exist within governance (e.g., AI oversight boards and algorithmic impact assessments) but commonly elicit participation from experts\footnote{As an example, consider  H.B. 1384, 33rd Leg., Reg. Sess. (Haw. 2025) and H.B. 3808, 89th Leg., Reg. Sess. (Tex. 2025)}. Further, language enacting public governance participation is commonly vague, or shifted to informally mean participation from quasi-representatives (e.g., civil society organizations) \cite{kaminski_stakeholder_2022}. The implicit self-interests of policy makers and experts (i.e., academia or private sector) can result in narrow conceptions of policy creation and can ignore marginalized experiences \cite{Mansbridge_Bohman_Chambers_Christiano_Fung_Parkinson_Thompson_Warren_2012}.

In this paper, we use collaborative autoethnography \cite{lapadat2017ethics, hughes2016autoethnography} to present a case study examining the benefits and challenges of using participatory methods to craft and shape AI policy. Our discussion of participatory AI governance methods is aligned with works on participatory governance \cite{maas2024beyond, kaminski_stakeholder_2022} not to be confused with participatory AI \cite{corbett2023power,birhane2022power}. Where the latter focuses on increasing public participation in AI design and development, the former considers public participation for public policy design.

Motivated by a desire to mitigate harm toward queer people from AI usage, our organization,
focused on raising awareness of LGBTQIA+ issues in AI and machine learning and fostering a supportive and decentralized community of queer researchers, began engaging in AI policy. Our organization's policy approach was shaped by two key factors: (i) organizational structure and (ii) organizational research on inclusive design and application of AI systems.

We present the following contributions detailing how our attempts at implementing participatory methods --- both within our organization and in broader policy processes we partook in --- succeeded and failed:

\begin{enumerate}[label=(\roman*)]
    \item We present a case study of participatory design (PD) in the development of a queer-focused US AI policy explainer surfacing challenges and solutions (in \S \ref{subsec:org_approach_to_pd}).
    \item We describe challenges to implementing participatory methods in policymaking in our involvement with NAIAC and NIST (in \S \ref{subsec:lar_sca_part}).
    \item We provide suggestions to policy makers to increase public participation from marginalized communities (in \S \ref{subsec:sug_to_pol}).
    \item We provide other civil society organizations working in technology and AI spaces with lessons and reflections to aid in their advocacy efforts (in \S \ref{subsec:sugg_to_orgs}).
\end{enumerate} 

\section{Background}

In this section we provide an overview of recent developments in US AI policy, then describe contemporary debates within PD.

\subsection{US Federal AI Policy: An Abbreviated History}

In late 2022, the White House Office of Science and Technology Policy (OSTP) released the Blueprint for an AI Bill of Rights \cite{OSTP_AIBillofRights_2022}, detailing broad principles for safe, effective AI use, development, and deployment. Its introduction was a federal signal for comprehensive AI governance goals, and was quickly followed by several proposed congressional bills \cite{brennan_center_2025}.

Biden EO 14110  \cite{biden_14110}, was introduced in late 2023 to consolidate AI governance conversations and announce long-term goals for the administration. It also precipitated the establishment of the National Institute for Standards and Technology (NIST) AI Safety Institute (AISI) and Consortium (AISIC)\footnote{NIST AISI was renamed the Center for AI Standards and Innovation in 2025}.  Biden Executive Order (EO) 13988 \cite{biden_13988}, was introduced in early 2021 and instructed executive agencies to extend Title VII of the 1964 Civil Rights Act to include sexual orientation and gender identity as protected classes. Biden EO 14110 \cite{biden_14110} was introduced after the Blueprint for an AI Bill of Rights \cite{OSTP_AIBillofRights_2022} by the Office of Science and Technology Policy (OSTP), which broadly defined individual data rights that Americans should maintain. These documents show a combined policy approach within the Biden administration between civil and data rights, evidence-based frameworks, and risk-management based approaches \cite{OSTP_AIBillofRights_2022}. More recently, are the Trump Administration's EO 14179 on an ``AI Action Plan'' \cite{trump_14179,trump_ai_action_2025} and US remarks at the Paris AI Summit\footnote{https://www.elysee.fr/en/sommet-pour-l-action-sur-l-ia} which marked a shift away from regulation towards AI as a tool for competing with other nations. 

Among the policy styles popular in Biden-era, rights-based policy approaches advocated for explicit protections on usage and types of acceptable personal data for use by predictive algorithms \cite{OSTP_AIBillofRights_2022}. Risk-based policy approaches, advocated for protections on the basis of harm categorization where higher risk systems would receive stricter protections and lower-risk systems lower regulation \cite{NIST_us_artificial_2024}. A wider coalition in academia and government advocated for evidence-based approaches \cite{Casper2025PitfallsOE, bommasani2024path, newsom2024sb1047}. Supporters of evidence-based approaches desired that empirical evidence be shown regarding AI risk and harm before suggesting policy considerations and stood in opposition to ``existential-risk'' by many in the AI Safety community \cite{statement_on_ai_2024}. References to each of these approaches appear in the NIST AI Risk Mangement Framework \cite{NIST_us_artificial_2024}.

The creation of the NIST AI Advisory Council (NAIAC) \cite{hr6216_2020} and AI Safety Institute (AISI) and Consortium (AISIC) \cite{aisic_2023} focused on responding to EO 14110 and codifying risk within the NIST AI Risk Management Framework (RMF) \cite{NIST_us_artificial_2024}. While many AI governance mandates during the Biden administration explicitly applied to executive agencies, there was an implicit expectation of wider adoption by private entities. Regarding background, NIST AISIC members came from academia, civil society, and private companies \cite{plannational}. Members participated in working groups designed to codify harms, socio-technical impact, and risk via red-teaming activities (i.e., empirical risk exploration). AISIC initially pushed socio-technical inclusion through a single working group; however, by the end of 2024, this group had been removed, signaling a shift a towards purely technical interests \cite{Oduro2025TroublingTS}. 

% In crafting policy directions, we had discussions with long established civil society organizations and technology policy experts.

\subsection{Participatory Design: A Primer}
Participatory design (PD) can be traced back to the 1970s workplace democracy movement in Scandinavian countries, with the fundamental \emph{political} ideal of emancipation by shifting and reclaiming power by those who lack it \cite[\S 2]{bodker_participatory_2018} --- a constant refrain as PD enters into its fourth era now with special emphasis on feminist, queer, indigenous and/or decolonial frames, among others.
Reflecting on the history of PD, Kensing et al. \cite{kensing_heritage_2013} proposed six guiding principles: \textit{equalizing power relations}, \textit{democratic practices}, \textit{situation-based actions}, \textit{mutual learning}, \textit{tools and techniques}, and \textit{alternative visions}. Similarly, B\o dker et al. \cite{bodker_participatory_2022} highlight the following four strong commitments in PD (\S 2.1, p.\ 7):
\begin{enumerate}[label=(\alph*)]
    \item \emph{democracy} at the workplace and beyond;
    \item \emph{empowerment} of people through the \emph{processes of design};
    \item \emph{emancipatory practices} rooted in mutual learning between designers and people; and
    \item seeing humans as \emph{skillful} and \emph{resourceful} in the development of their future practices.
\end{enumerate}

Additionally, it is helpful to introduce the roles of \emph{user} and \emph{designer} common in PD language \cite{Velden2015ParticipatoryDA,Bdker2010ParticipatoryDI}. Traditionally, there exists a unidirectional designer-user relationship with power flowing from designer to user (i.e., users have little influence over how the system is designed). PD attempts to shift this power relationship towards co-directional designing and development.

\paragraph{PD in governance}Likewise, participation in public governance benefits from diverse view points. As an example to the ills, a 2013 Seattle City Police Department secretly acquired a surveillance drone and mesh network (to facilitate surveillance cameras), which effectively bypassed relevant city ordinances. Only after it was met with large public disapproval \cite{Clarridge_2013a} were both programs terminated \cite{Clarridge_2013b}. In this instance, public participation provided oversight --- expanding harm considerations in city governance and highlighting technical translation discrepancies among stakeholders \cite{crump_surveillance_2016,doi:10.1177/2053951719868492}. However, the public's direct participation in AI governance is challenging, as policy advancements are a non-linear combination of informal interactions, timing, and people, creating a ``perfect storm'' where policy advancements occur only under certain conditions.

\paragraph{Scaling conflicts} Scale can complicate PD in practice. Here we follow \citeauthor{bossen_scaling_2025} to understand scaling by its temporal, socio-spatial, and onto-epistemological characteristics \cite{bossen_scaling_2025}. In the same context, scholars have also spoken of the constitutive nature of ``infrastructuring'' in scaled PD, or adaption of smaller co-design infrastructures for sustainable, large-scale PD projects \cite{Neumann1996MakingIT,karasti_infrastructuring_2014}. An important account is given in \citet{bodker_tying_2017} where the notion of a spectrum between relatively stable \textit{networks} and more fluid and transient \textit{knotworks} can be mobilized to illustrate the importance of backstage activities in participatory infrastructure. The latter considers both formal and informal aspects of coalition building and inter-team relationships. In this paper, we present our involvement in NIST AISIC and creation of a queer-focused US policy explainer as case studies to consider the temporal, socio-spatial, and onto-epistemological scaling challenges of using PD principles in practice.

\section{Engaging in AI Policy as a Grassroots Organization}

\begin{figure}
    \centering
    \caption{A timeline of Queer in AI's participation in US AI governance efforts.}
    \includegraphics[width=\linewidth]{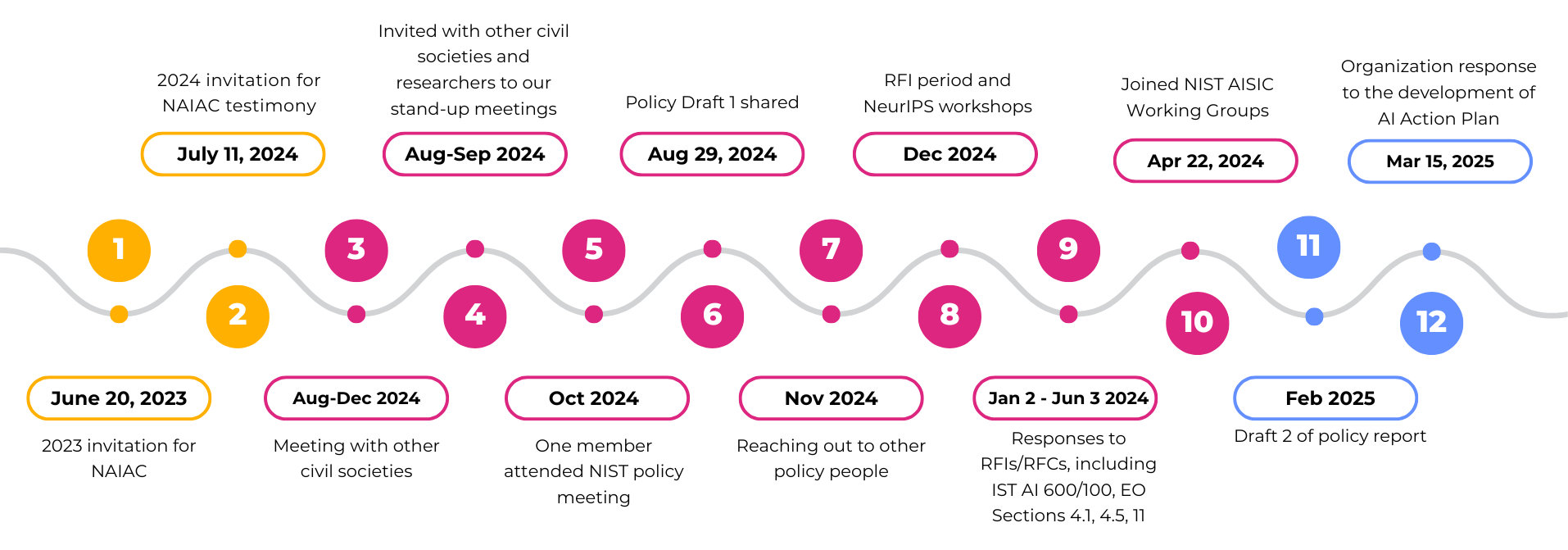}
    \label{fig:participation_timeline}
    \vspace{-1cm}
\end{figure}

\subsection{About Queer in AI}

% \begin{figure}
%     \centering
%     \begin{tabular}{ | c | c | }
%     \hline
%  P1-P7 & PhD students in computer science  \\ 
%  \hline
%  P8 & Postdoc in computer science  \\  
%  \hline
%  P9-P13 & Professionals working in software \\
%  \hline
%  P14 & Assistant Professor \\
%  \hline
%  P15 & On job market
%  \hline
% \end{tabular}
%     \caption{Overview of some quoted participants in our policy work}
%     \label{tab:participants}
% \end{figure}

Our organization represents both queer people working in AI and related fields, and the needs and interests of queer people broadly in AI. We advocate for effective Diversity, Equity and Inclusion (DEI) policies, workplace protection, and other measures to ensure queer people enter, stay in, and have fulfilling careers in AI. Additionally, we advocate for AI, data, and technology that respects privacy \cite{shan2020fawkes}, counters surveillance \cite{cherepanova2021lowkey}, represents fluid and complex queer identities \cite{QueerInAI2023BoundBT, Haimson2025TransT}, and does not degrade the environment or undermine creative economies. Finally, we fight for positive uses of these technologies for queer people, particularly in protecting and improving the online spaces where many queer people find community and knowledge.

We use a flat organizational structure, with only three formal roles: \textit{members}, those who are in the virtual workspace or email list; \textit{organizers}, those who are or have organized an event or initiative; \textit{core organizers}, people who have been organizers for several years. Our organization largely functions on the platform of a Slack workspace, where almost all channels are open to everyone with the exception of a channel where core organizers infrequently discuss issues involving personally identifiable information (PII) or topics with a high chance of negatively impacting individuals or the organization should they be widely publicized. 

%In Figure \ref{fig:participation_timeline} we provide a timeline of our policy work. 
Internal conversations regarding queer perspectives in AI governance began in 2022 following Biden's Executive Order 13988  on preventing discrimination based on gender identity or sexual orientation \cite{biden_13988} and our invitation to testify before the White House National AI Advisory Committee in 2023. Formal engagement in US AI Policy began in 2023 following Biden's Executive Order 14110 on safe, secure and trustworthy AI \cite{biden_14110} and the resultant creation of the NIST AI Safety Institute (AISI) and Consortium (AISIC). Both executive orders had far reaching effects for anyone accepting federal funding and were also accompanied by a toolkit to codify equality efforts for transgender and LGBT+ individuals \cite{whitehouse_trans_toolkit}. The scope of EO 14110 was noticeably wide but directed the work of two groups, the NIST AI Safety Institute (NIST AISI), and the AI Safety Institute Consortium (AISIC), in producing policy guidance for executive agencies and those receiving executive funding. 
Our organization desired to reflect findings from our experience in community engagement and inclusive AI research in policy conversations.

In 2023, a policy action group in response to increasing requests for our organization to take on policy positions was launched. The policy action group existed as a public Slack channel, a weekly stand-up meeting, and several in person events. Both the Slack channel and the stand-up meetings were open to all members, leading many members to join for the duration of the initiative. The policy group was dedicated to understanding what engagements, research, and documents would advance the needs of queer people in AI policy, and pursuing those engagements. The policy group also produced a queer-focused US AI policy explainer, collecting our community's positions and relevant research.

\subsection{An Organizational Approach to Increasing Participation}\label{subsec:org_approach_to_pd}

\begin{figure}
    \centering
     \caption{Diagram illustrating power flow between ``Users'' (policy groups and the general public) and ``Designers'' (government entities). Connections between Users and Designers demonstrate the specific ways each group influences AI policy. }
    \includegraphics[width=0.8\linewidth]{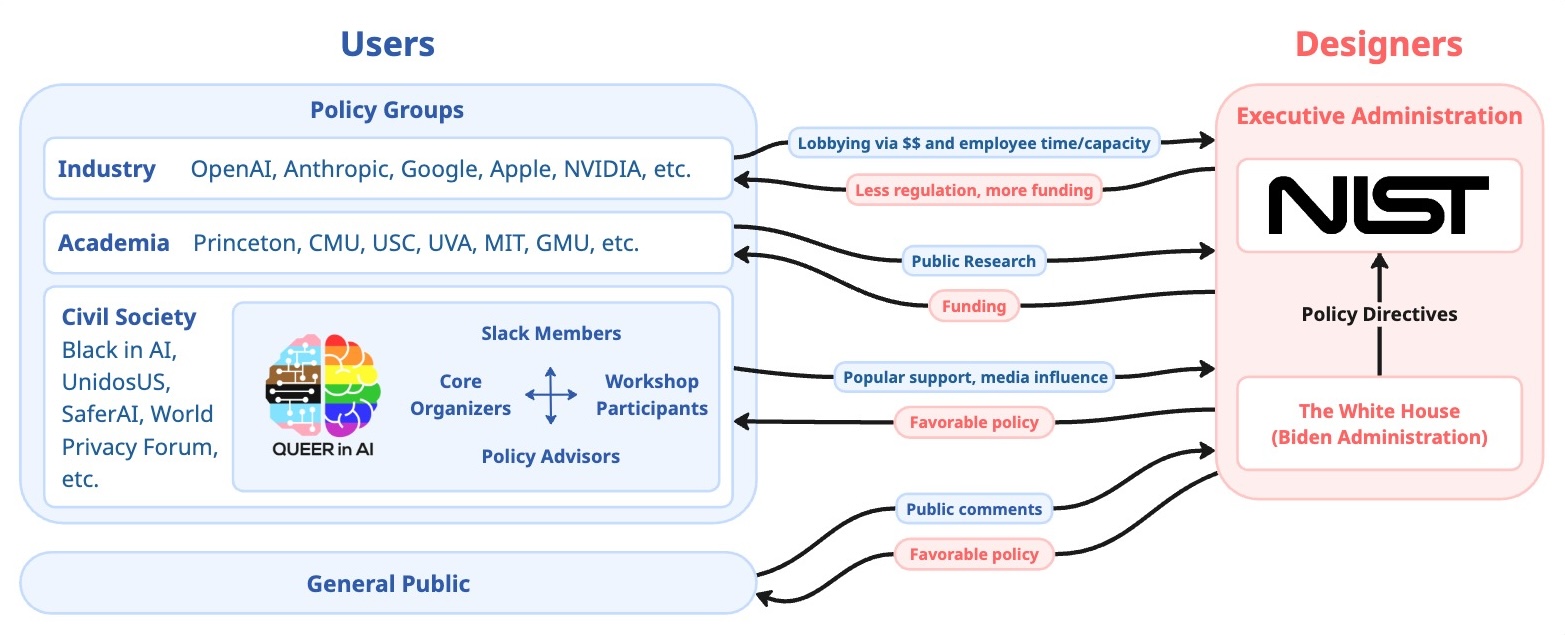}
    \label{fig:usersDesigner}
\end{figure}
The process of describing our work within a larger context is autoethnographic in nature. Autoethnography as a practice analyzes personal experience (i.e., \textit{shares characteristics of autobiography and ethnography...} \cite{Ellis2010AutoethnographyAO}). At its core, autoethnography centers on ``extracting meaning from experience...'' \cite{bochner_criteria_2000}  within a larger social and cultural context \cite{Ellis2003TheEI,Adams2008AutoethnographyIQ}. Methodologically, autoethnography may use a mix of lived experience, retrospective interviews, text, and tools both familiar to the autoethnographer and not. As such, our collaborative autoethnographic practice is informed by our varied research backgrounds and organizational interests in participatory AI.

We adopt the PD principles defined in \citet{kensing_heritage_2013} to guide discussion for how participatory design (PD) practices were used in our organization's structuring and design of a policy explainer. Our organization takes on the role of \emph{designer}, when deciding organizational policy direction, and \emph{user} when participating with NIST AISIC. We note the presence (or absence) of PD practices for both.

\paragraph{Equalizing power relations} Across PD is a goal of placing power closer to people and communities \cite{Ehn1992ScandinavianDO,birhane2022power,smith_routledge_2024}. In both governance creation and AI development, there commonly exists an initial communication barrier created by the lack of domain knowledge (e.g., policy and architecture design). Organizations practicing PD may first incorporate training or educational components before attempting system co-design with users.\footnote{As an example consider technology literacy initiatives run by the \href{https://detroitcommunitytech.org/?q=projects}{Detroit Community Technology Project}.} While participation in governance traditionally necessitates specialized skill (i.e., knowledge of laws, legislation, key entities), by widening the definition of skill to include lived-experience, we importantly expand \textit{who} can use the resultant system. More traditional participation channels between NIST and consortium members included regular meetings (known as ``office hours'') and requests for comment or information (RFC and RFI, respectively) in response to the deliverables mentioned in Biden EO 14110 \cite{biden_14110}. As an organization, we considered open access to information, online meetings, and knowledge sharing (i.e., of queer and marginalized lived-experiences, policy and technical experience) to minimize dominating factors and encourage broad participation.

\paragraph{Democratic participation practices} In NIST AISIC, democratic participation existed most commonly in the research directions chosen in the sub-working groups. NIST AISIC was also a quasi-open entity, consisting of explicitly invited groups and those who applied to join during open invitation periods. While a large number of groups joined, fewer were active in the projects across working groups and in contributing to key documents like the AI Risk Management Framework  (RMF) \cite{NIST_us_artificial_2024}. As an organization, we allow open membership through word of mouth and outreach at AI conferences. Within our policy working group, we decided our policy stances collectively within these open meetings, allowing all members to contribute to organizational policy. These policy stances were also brought to a wider AI research community in conference workshops, compared against policy stances taken by other civil society groups (i.e., focused on equitable tech for marginalized groups, tech justice, mitigation of representative harms), and evaluated against the lived-experiences of those within the organization. We advertised public requests for comment within our Slack community and publicly on our website. We aligned our policy stances to the US general queer public though a combination of litmus tests -- engaging advisors, others in the policy community, civil society groups, and news related to queer representation and technology.  Organizational transparency was meant to lower barriers to participation and encourage members to become more involved in organizing. We also incorporated yearly feedback from the larger AI community at AI conferences. We illustrate this iterative process in Figure \ref{fig:policy_design_flowchart}.

\paragraph{Situation-based actions} In PD, it is necessary to involve those directly affected by the system being designed \cite{smith_routledge_2024}. In US AI governance, this necessitates understanding the broad lived experiences of citizens globally. At smaller scales, this is accomplished by working directly with system users. While queer people represent a small portion of the general population, it is still intractable to consider the lived experiences of all queer people across the US. NIST AISIC and our organization both operate as quasi-representatives of the general public. This realization introduces bias in priorities, knowledge, morals, and other contextual differences not shared by the public.

\paragraph{Mutual learning} In our organizational role as \emph{designer} (of policies representative of queer communities), we have limited formal chances for mutual learning with the queer US general public. We instead shift into a role of \emph{user} and engage in discussion with and mentorship from other grassroots and sociotechnical organizations. This both improves our collection and outreach practices, and attempts to align our policy projects with community needs. Within distributed PD \cite{Obendorf2009InterContextualDP, WINSCHIERSTHEOPHILUS2022100439} are considerations for how to represent multicultural and multicontextual experiences. Though we shared a common set of principles towards inclusive AI development and policy creation we came from different areas of practice (i.e., in research, academia, industry, etc.), different experience levels, geographic locations, and cultural backgrounds. Thus, mutual learning could take place in those areas --- in smaller breakout meetings, through invited speakers, and informally in weekly meetings. We also pushed for mutual learning during retrospectives with the larger AI and queer community (e.g., 2024 NeurIPS workshop)

\paragraph{Tools and Techniques} Within NIST AISIC, tools such as Slack, email, and online and in-person meetings allowed for broad participation. There appeared a mix of formally scheduled meetings and informal break outs (i.e., in response to subgroup objectives). Our policy efforts and activities also depended heavily on both technological and social infrastructures for distributed work via virtual platforms such as Slack, Zoom and Google Docs. These tools facilitated collaboration across diverse geographic locations and timezones, and with other advocacy groups with similar visions of AI governance --- we were thus able to build \emph{collectives} with those groups. Importantly, we crafted an organizational policy explainer as a tool to translate research and lived experience to policy makers and groups. Members naturally used methods common to their subfields (e.g., research methodology, visualization methods) allowing for translation between research and policy audiences. Also of note is our use of the Chatham House Rule\footnote{\url{https://www.chathamhouse.org/about-us/chatham-house-rule}} in conference workshops and reflective breakout sessions, which attributed discussion to the group as a whole and not the originating party.

% We regularly seek feedback from outside of our group through public comment periods and targeted feedback periods. We maintain best practice through mentorship from established policy organizers and groups and engaging in group research and discussion on community engagement and participatory methods (cf.\ \citealp{qinai_2023_case_study, oduro_ai_2024, bogen2024}). 

\paragraph{Alternate Visions} The important arrival point from PD is that of alternative, meaningful system solutions, especially ones that challenge existing norms \cite{zahlsen_challenges_2023}. The origins of AISIC from EO 14110 \cite{biden_14110} limited opportunities for democratic solutions and co-design (i.e., timing and scope). Scale also introduces conflict at many points in the system co-design process \cite{zahlsen_challenges_2023}. Within AISIC, this conflict existed as members had varying levels of familiarity with projects (i.e., due to existing relationships with NIST, geographic location to other members and Washington D.C.) outside of the stated project goals. Organizationally, alternative system design existed in how we incorporated ontological differences into our work, by ``accepting people in context'' \cite{simonsen_routledge_2012}. As an organization that originated in the US, we also explicitly considered how to expand our programming and projects beyond this area (i.e., in changing communication style, where we advertise, where we recruit new members, accessible meeting times, etc.). Smaller working groups could adopt different communication styles and iterate over ideas together, away from the norms common to the larger group.

\subsection{Involvement with NIST}

In 2023 and 2024, our organization was invited by members of the White House National AI Advisory Committee (NAIAC) to give testimony on harms caused by AI systems and related entities (e.g., company data use policies, scope of pre and post deployment evaluation) to queer people. Our group's initial 2023 testimony was given along with other, similarly focused, research-based, AI affinity organizations and civil society organizations. The 2023 session allowed each group an initial two-minute statement and limited discussion after. The 2024 session allowed five minutes for an initial statement and a longer discussion. Both sessions were closed, only open to members of the committee, and NAIAC committee attendance appeared optional. Some of our organization’s appeals do appear in year 1 and 2 reports \cite{naiac_2023, naiac_2024} and resultant recommendations to NIST AI Risk Management Framework (RMF) \cite{NIST_us_artificial_2024}. However, we saw little concrete impacts result from our participation in the NAIAC, with one of our members noting, "It did not feel like the NAIAC sought longer-term engagement with the concerns that we raised in our testimony beyond the listening session."

%There were also discussions with researchers who had held government-appointed policy positions, further validating these stances and speaking to their clarity for policy audiences. NAIAC briefings were to an audience composed primarily of academia, industry, and civil society. Many committee and consortium members had some legal expertise. As our expertise largely was not in law, we frequently needed to translate our research to agency specific applications and cognizable claims.

% As the only queer focused group to give testimony to NAIAC, the scope of our testimonies often did not show a clear plan for describing the needs of queer people in the US. The practical reason for this is that this exceeded the scope of what our volunteer policy team could accomplish. We believe more concrete plans for our organization to facilitate larger engagement should have been considered.

Our organization then applied to join the NIST AI Safety Institute (AISI) and Consortium (AISIC), aiming to advocate for some of the calls in our 2023 and 2024 testimonies to NAIAC. 
Many organizational meetings were held to gather initial thoughts and develop strategies focused on advancing rights-based governance strategies. As the only queer-focused organization, the group decided to be vocal about queer-specific problems such as forced ``outing” of queer people in unsafe or malicious contexts, misgendering, consent, and excessive or insufficient data collection and use. To further amplify our points, our organization partnered with more established policy groups and larger queer-focused organizations outside of AI and technology spaces. These meetings considered how to reflect the classification of harms from ethics research \cite{Slattery2024TheAR} to NIST RMF \cite{NIST_us_artificial_2024}, which chose to focus on risks specific to generative AI (e.g., bio-warfare and explicit child sexual abuse material in LLMs). The NIST RMF did not explicitly focus on wider issues of model bias and discrimination present in generative and non-generative systems. Thus, dominant conversations within NIST AISIC focused on narrow definitions of AI systems and related harms with limited integration of bodies of knowledge built by the AI ethics community. This realization was reflected in formal NIST meetings, responses for comment or information, and with other civil society groups within AISIC. Our organization noted that their involvement with NIST felt ``opaque” at many points. We did however, continue to submit responses to agency requests for information.

\subsection{Policy Explainer}
\begin{figure}
    \centering
    \caption{An illustrative diagram of the design process for our organization's policy explainer. Two full collaborative iterations were performed. Date ranges are shown for both design sessions. }
    \includegraphics[width=\linewidth,trim={0 0 0 5cm},clip]{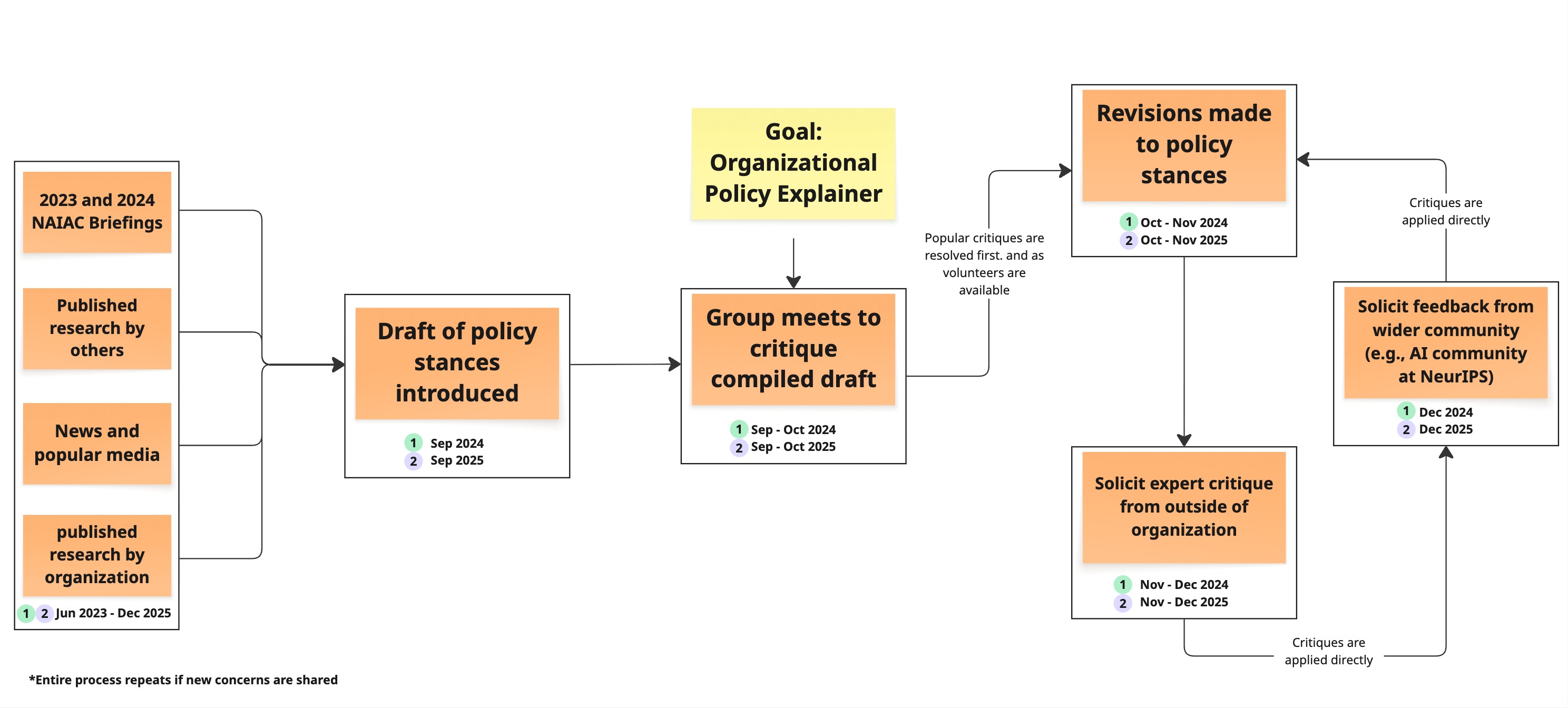}
    \label{fig:policy_design_flowchart}
    \vspace{-0.9cm}
\end{figure}
%
% Initial draft of explainer
The centerpiece of our PD efforts was the creation of a queer AI policy explainer (cf.\ Fig~\ref{fig:policy_design_flowchart}). In September 2024, our organization started the process of drafting the explainer, with the first draft outlining the risks that AI systems posed to queer and other marginalized communities, highlighting historical and present harms such as discrimination, violence, erasure, and pseudo-scientific uses (e.g.\ automatic gender recognition), while also calling for greater transparency, consent, and privacy protections. This first draft was then distributed to participants at a NeurIPS workshop in December 2024. Participants read the draft in two sessions, each divided into groups focusing on one section of the explainer. Comments from participants were collected per group, each written on a sheet of paper and transcribed by an organizer. We received detailed critiques on the contents of the draft from the participants: One raised the ``need [of] a more refined definition of `outing' [and of] `doxxing' [...] to include wider aspects of queer identity[.]'' On the issue of pseudoscientific use of AI, one commented that ``[p]seudoscientific AI feels [...] ambiguous compared to AI to support or legitimize pseudoscientific purposes \textcolor{red}{that cause \underline{harm}}\footnote{Color and underline in original}''. Others suggested that the draft needed more discussion on the ``[h]omogenization, lack of community perspective, erasure'', ``[i]ntentional connections get[ting] loss[\emph{sic}], [and] loss of community''.
\iffalse
% Transcription with rich formatting
``Pseudoscientific AI feels \st{less} ambiguous compared to AI \st{for [...?]} to support or legitimize pseudoscientific purposes \textcolor{red}{that cause \underline{harm}}''
\fi
%
Following the workshop, we extensively modified our policy explainer draft to take feedback into account: most importantly, we shifted away from trying to address any specific regulation or legislation in any jurisdictions to tailoring the explainer specifically to the US context as well as articulating a set of principles for AI policy that took the needs of queer people into account in order to create a more general and clearer document\footnote{The current version of our policy explainer is available at \url{https://www.queerinai.com/policy-research}}.
%
% Eliciting feedback at NeurIPS 2024 workshop
% What did people like?
% What critiques did people have?
% How did we change explainer to respond to this?
% What did we do well and not well in getting this feedback?
% 2-3 anonymous participant quotes to illustrate these points

% We felt we were engaging consistently and critically but without metrics gauge whether our concerns were being considered. This can be considered a shortcoming of the existing policy communication structures with executive agencies, of which our organization may not have been fully aware. Our organization met with AISIC leadership and discovered that, while their participation had been consistent, our comments had not nessecarily been read by current NIST staff.

\section{Participatory Methods in Policymaking: Retrospectives}\label{subsec:lar_sca_part}

\begin{table}[]
\renewcommand{\labelenumi}{\arabic{enumi}.} %swtich numbering back
    \caption{Partial table of organizational challenges to using PD and our approaches for mediating}
    \centering
    \footnotesize
    \begin{tabular}{p{7cm}p{8cm}}
      \hline
\textbf{Organizational Challenges} & \textbf{Organizational Approach }\\
\hline
AI policy work is often in-person in a few select geographical centers. & 1. Partner with like-minded organizations who can participate in person\\
      \hline
      AI policy is often opaque. & \vspace{-0.2cm}\begin{enumerate}[leftmargin=*]
            \item Mentorship from more experienced organizations.
            \item Coalition building with other organizations with more ``access''
        \end{enumerate}\\
      \hline
      Only queer-focused and one of few civil society organizations,
        which created high expectations,
        pressure, and feelings of exploitation. & \vspace{-0.2cm}\begin{enumerate}[leftmargin=*]
            \item Coalition building with other organizations to raise this point to policy leaders
        \end{enumerate}\\
        \hline
        
        Members had high workloads from other academic \& industry positions, limiting participation in unpaid policy work.& \vspace{-0.2cm}\begin{enumerate}[leftmargin=*]
          \item Increase the pool of volunteers to allow enough active membership as availability shifts
          \item Use open meeting structure and web-based documents for asynchronous work
        \end{enumerate}\\
        \hline
        
         AI, data, and digital technology have
        vast impacts on queer people, making
        the scope of focus challenging. Few groups are tasked with addressing
        the needs of many, diverse communities.& \vspace{-0.2cm}\begin{enumerate}[leftmargin=*]
            \item Coalition building with other organizations to increase the scope of issues covered
            \item Elicit more direct feedback at NeurIPS workshop
        \end{enumerate}\\
        \hline
        
         Formal communication structures allow
        limited stakeholder participation& \vspace{-0.2cm}\begin{enumerate}[leftmargin=*]
            \item Policy decisions factor in organizational feedback from a demographic survey 
            \item Elicit more direct feedback at NeurIPS workshop
            \item Use open meeting structure
        \end{enumerate}\\
        \hline

        Existing mechanisms for public
        participation (e.g., Requests for
        Information/Comment) are not suited
        for layperson participation. & \vspace{-0.2cm}\begin{enumerate}[leftmargin=*]
            \item We operate as a quasi-representative for the queer public as many members are queer and knowledgeable of AI and governance mechanisms.
            \item We ``translate'' needs from informal news and social media to relevant policy asks.
        \end{enumerate}\\
        
        \hline
    \end{tabular}
    \label{tab:organizational_challenges}
\end{table}

Public participation within the US executive administration is used commonly as an oversight mechanism -- a role separate from the equilateral co-design relationship this work explores. Public participation in the former can exist as public comments, documents, materials, and meetings. Within NIST AISIC, the public could apply to participate (with final invitation decided by NIST), all document creation and comments were public, and internal meetings between NIST AISI and consortium members were private. The level of public participation was decided solely by NIST (within reason of transparency regulations such as the Freedom of Information Act \cite{foia_1967}) and included limited avenues for public participation and collaboration (i.e., collaboration in writing NIST AI RMF \cite{NIST_us_artificial_2024}.

\paragraph{Participatory methods struggle to scale across long distances and handle sudden spikes in workloads}  Our organization experienced socio-spatial and temporal PD scaling conflicts surrounding work in the US policy space. Within NIST, key documents where public response was essential \cite{NIST_us_artificial_2024, nist_a_plan_2024,nist_secure_software_2024} had small response windows and appeared within a small timeframe. This created sudden spikes in labor needs, affecting volunteer and decentralized organizations negatively. Additionally, it may have biased responses towards organizations with prepared knowledge bases over those concurrently gathering community feedback, groups synthesizing relevant research, and decentralized groups working across timezones. One way our organization mitigated this challenge was by creating a policy explainer. Any member could refer to the information in this document instead of the varied documents and discussions behind it's creation. Direct participation with queer people across many contextual backgrounds is ideal, but creates a large resource overhead, since our members live in many different timezones and cannot all meet at one time, much less travel to one place. The overhead of many sessions to compile information from smaller, more homogeneous teams (e.g., by geographic region or interest) within the policy working group presented a roadblock instead pushing us to compile information based on resource and time availability (e.g., once yearly and in-person at conferences). Additionally, technological requirements (e.g.\ internet access on compatible devices)  for participating in our policy efforts were biased against queer folks residing in the Global South. Fully mitigating geographic biases was beyond the scope of our work though, it was still necessary to eliminate immediate conflicts preventing wide organizational participation. As an example, hosting virtual meetings on different platforms and practicing information redundancy (i.e., sharing relevant information across multiple platforms) alleviated barriers such as IP blacklisting or time zone differences but could not fully eliminate them. In cases where someone had limited network or technology accessibility, our efforts would have little effect. Finally, associated with scaled PD are diversified onto-epistemologies, necessitating many parallel points of feedback, contribution, and co-design in support of a single project \cite{1579648, bossen_scaling_2025,roland_p_2017}. Organizationally, we used feedback from the entire organization (through demographic surveys) to offer an additional source of contribution. By creating a centralized policy document containing our collective thinking and prioritizing many different methods of participating outside of meetings, our organization was able to address challenges of scaling PD across long distances and to short timeframes.

\paragraph{Organizational conflict in scaling PD} Associated with scaling PD are myriad conflicts. Involving many members in policy co-design discussions was complicated by timezones and other constraints preventing everyone from meeting at once. Scale was instead accommodated by the regular rotations in participation from members (based on academic schedules, busy work periods, etc.). Members could participate on projects asynchronously via Slack and web-based shared documents. New members to the working group appeared to naturally defer to more senior academic members, echoing power dynamics common in US academia. Several new members felt a hesitance to conflict or disagreement on the basis that some members ``knew more.'' More experienced members in an organization may naturally feel more ownership in system design. Knowledge sharing between newer and more experienced members via mentorship can facilitate co-ownership. Collective knowledge development both aid planning and engender agency \cite{Freire2019PedagogyOT}. In an effort to equalize power relationships our flat organizational structure, open meetings, and opportunities for mutual learning had some positive effect, but ultimately more avenues were needed. As an example, we would have benefited from mutual learning paradigms outside of those common in academia (e.g., interactions with other academics, familiarity with research literature, paper writing). While these communication styles were familiar to many in the policy group, they were not fully observant of the diverse contextual backgrounds of those in our group. Socio-spatial diversity also introduced conflict without inherent pathways to democratic solutions. Our disagreements sometimes appeared due to varied value systems (e.g., varied techno-political values). In many cases, the loudest, most consistent voice was adopted and in other cases, disagreement spurred another iteration of larger critique (seen in Figure \ref{fig:policy_design_flowchart}). Thus PD-based solutions sometimes created an ongoing loop of discussion where solutions died due to timing or resource limitations. In other cases, we asked the originating member to lead crafting a solution with others, effectively forming new projects.

\paragraph{Lowering barriers to participation} Our organization established a large, active Slack space with over 1,600 members, and by repeatedly advertising the policy team and its meetings, particularly when members organically brought up policy concerns, we were able to recruit many volunteers. Barriers to joining the Slack channel and weekly meetings were also minimized; the slack channel was kept public, and anyone could join the meetings. We also hosted a policy workshop, co-located at NeurIPS, where a broader set of community members had a chance to engage with our organization, join our Slack channel, and connect with our policy team. Many organizers noted that this openness was in unresolved tension with another key value of online queer spaces: privacy and security. Organizers worried that the openness of the policy team could lead to malicious actors joining and disrupting proceedings or doxxing participants, mirroring frequent concern for online queer communities \cite{meta_2024}.

\paragraph{Industry capture of regulatory processes} 
In AISIC, civil society and other policy groups supporting marginalized communities, reported feeling that their organizations were sidelined in NIST efforts, weakening feelings of co-ownership. Supporters of equitable AI, civil society, and other policy groups frequently referenced the membership of AISIC,\footnote{\url{https://www.nist.gov/aisi/artificial-intelligence-safety-institute-consortium/aisic-members}} where for-profit corporations comprised the majority of the membership, the industry-heavy compositions of high profile AI policy events \cite{readout_2024}, and the industry backgrounds of key AI appointees \cite{Primack-MAGA-Outrage} as evidence for this belief. Our organization submitted multiple responses to NIST RFIs, only to find out that those responses had not been read or taken into account when we asked NAIAC organizers. Our organizers noted that NIST seemed eager to form deep partnerships with industry AI developers \cite{Press-NIST-Testing}, but the role of civil society was much less clearly defined.

\paragraph{AI policy work is often in-person in few select geographical centers} Although our membership and volunteers are spread throughout the world, with particular concentrations of members in the United States, no member of the policy team was based in cities hosting major AI policymaking institutions or events, most often Washington D.C., London, Brussels, or San Francisco. Our organization could not afford to hire full-time staff, or pay for policy team members to travel to AI policy events, which prevented us from attending almost all AI summits, workshops and other events, and crucially from engaging in activities surrounding such events -- dinners, side meetings, other networking -- that could have helped us meet allies, improve our knowledge of what was going on, and build our organization's ``inside game'' \cite{insideoutside} (cf.\ \citealp{bodker_tying_2017}).

\section{Discussion}

\subsection{Suggestions to Policy Makers}\label{subsec:sug_to_pol}

\paragraph{Opaque relationships hurt effective AI regulation}
We observed that civil society groups were already disempowered entering into AISIC, and we believe that coming together and building collective power is critical to addressing the vast interests opposing effective AI regulation. Nonetheless, we frequently heard about other meetings and groups similar to ours, generally all opaque, secret, and invite-only. While there are reasons for some discussions to be private, the default opacity in AI policymaking weakened collaboration between government agencies, people, and organizations. While our organization contained specialized knowledge as AI practitioners, many in the policy working group lacked knowledge inherent to federal-level policy creation, communication norms, and, \textit{critically}, knowledge of existing power flows to effectively engage with NIST. As a member of NIST AISIC, there was the assumption of participation in contributing to national AI standards. Additionally, explicit power flows existed between the Biden White House, NIST, executive agencies, and AISIC members as demonstrated in Figure~\ref{fig:usersDesigner}.

\paragraph{Limit extractive user-designer relationships by diversifying AI advisory councils} In NIST, our organizers worried about why policymakers were including our organization in meetings and programs, citing concerns that they were included to superficially boost the diversity of participants \emph{without challenging the underlying power structures} and that our positions and expertise had not been taken seriously. These concerns also mirror ongoing discussions of the pitfalls of participation within AI ethics literature \cite{Sloane2022ParticipationWashing, birhane2022power}. Our organizers continued to participate in policy efforts even when they had concerns about the intentions of engagement because they knew they were the only group invited to represent the needs of queer people, and refusing to participate would mean queer people would not have any voice. 
% added this bit
Increasing the diversity of marginalized groups in AI Advisory councils such as AISIC would work to mitigate these concerns and has seen support in recent state legislation\footnote{See HB 1916, 2025 Leg., 60th Sess. (Ok. 2025) and Ore. Exec. Order No. 23-26 (2025)}.

\paragraph{Limited meeting accessibility creates barriers to participation} All gatherings should be virtual or strong hybrid to increase accessibility and include public participation options.
AI policy work is often in-person in few select geographical centers. Although our membership and volunteers are spread throughout the world with particular concentrations of members in the United States, no member of the policy team were based in cities hosting major AI policymaking institutions or events, most often Washington D.C., London, Brussels, or San Francisco. Physical distance from important policy meetings and centers consistently excluded our organization from advocating for our positions, networking, and gathering information. Even if we were invited to every major AI policy event, we could not afford to send even one member of our team to participate in-person, nor could they take off that much time from their day jobs. Making all gatherings entirely virtual, or strong\footnote{Here \textit{strong} means that virtual participation does not play second fiddle to the in-person components and that adequate infrastructures (e.g.\ internet, virtual meeting, A/V) are in place} hybrid would greatly reduce this barrier. Greater public participation also prevents these events from becoming exclusive gatherings of elite academics, top government officials, and industry, as would providing public meeting notes or recordings. Lower resourced organizations suffer from a lack of full-time staff, pay incentive, and funding to travel to AI policy events, AI summits, workshops and other events. This also limits informal participation surrounding such events --- dinners, side meetings, and other networking --- necessary to build strong \emph{knotworks} across users and co-ownership in system co-design \cite{insideoutside, bodker_tying_2017}.

% Our policy initiatives represented a relatively successful attempt of scaling up PD socio-spatially and onto-epistemologically (cf. \citealp{bossen_scaling_2025}). 

\subsection{Suggestions to Organizations}
\label{subsec:sugg_to_orgs}
\paragraph{Power-imbalanced hierarchies are inherent among users in PD} Although PD is not inherently opposed to any hierarchy, there exists a tension between our stated ideal of decentralized organization and the reality that a few select core members, generally those having relevant domain expertise and occupying research positions, assumed the bulk of the effort and responsibilities. This suggests a need to reconsider organizational values or better ways to redistribute the authority and workload so that those \emph{core organizers} can function more like \emph{facilitators}. We also need to be vigilant to avoid extractive participatory research and design practices \cite{loi_participatory_2024}. Additionally, members introduce learned hierarchical relationships from other spaces, most commonly that of academia and industry seniority. Our organization's policy team members have been frequently multiply-marginalized, facing many barriers to their participation in this unpaid work; they were almost all students and junior academics, requiring each to balance coursework and pressure to publish against contributing their expertise to our organization's policy team. Some organizers reported institutional targeting for their activism and organizing work, pressure to simultaneously contribute to multiple organizations advocating for marginalized communities, illnesses, institutions and supervisors unsupportive of work in AI outside narrow technical domains.

% Most recently, within the United States, a changing funding landscape\footnote{https://www.americanprogress.org/article/mapping-federal-funding-cuts-to-us-colleges-and-universities/} has made it extremely difficult to support work focusing on AI and marginalized communities as concrete ways marginalization posed barrier to their policy work with our organizers. 

\paragraph{Organizational scarcity mindsets hinder policy advocacy} Mentorship by larger and more experienced organizations was vital to our efforts. However, obtaining this mentorship required years of building our organization's reputation and network, which in turn relied on many of our members privileged of being inside academia. Our members applied for many formal policy workshops, fellowships, and mentorship programs, but they were space limited and we were rejected from all of them. If we had relied solely on larger organizations or labs, our policy work would have never occurred. The scarcity model that many AI policy mentorship opportunities are built on can exclude marginalized communities.

\paragraph{Traditional communication structures limit user participation. } Current AI governance frameworks often fail to prioritize the concerns of marginalized communities \cite{parthasarathy2024bringing}. A common mechanism intended to address this gap is the Request for Information (RFI) process, used by organizations to solicit feedback from various stakeholders. RFIs are designed to democratize policy development, offering stakeholders, including marginalized communities, a platform to influence decisions that affect them. RFIs are an essential tool for fostering inclusive dialogue and ensuring that a wide range of perspectives are considered in the regulatory process. Research highlights that when RFIs function as intended, they can help incorporate a variety of voices, leading to more comprehensive and balanced policy outcomes \cite{Darbishire2010ProactiveTT}. In reality, the RFI process often falls short of achieving genuine inclusion. These mechanisms can be opaque and difficult to navigate, especially for marginalized communities that lack the resources or institutional knowledge to effectively participate. Furthermore, while RFIs invite public feedback, they often lack the transparency and accountability needed to ensure that this input leads to real policy change \cite{oneil2016weapons}. In a meeting with NIST AISI leadership, we were informed that our RFC/RFI responses hadn't been considered in their discussions. In NIST, our organization understood the communication norms to be through responses to NIST documents and participation in NIST AISIC activities though there appeared to be limited engagement with this work. The policy explainer was created to fill this gap, offering a clear, accessible framework to further empower public participation from marginalized groups in AI governance. Additionally, physical distance increased the difficulty of understanding relevant discussions and ideas within the design process. Our organizers often reported feeling confused and surprised by developments in AI policy. Specifically, members within our policy working group felt blindsided by releases of documents for public comments, changes in directions (or removal entirely) of working groups, and responses (or lack of response) to their feedback. Members discussed feeling uncertain about the purposes and intents of different policy documents: whether a bill was seriously intended to be passed, or just meant for signaling, whether a policy document would ever be meaningfully enforced. This uncertainty led organizers to be unsure of where to spend their limited time, to feelings that meaningful and critical conversations were had in meetings and spaces they were excluded from, and their efforts may have been wasted.

\paragraph{Greater participation in federal AI governance benefits from overlapping missions} Closely related to the breadth of issues we felt we needed to respond to is that we were typically the only queer-focused and one of few civil society organizations, which created high expectations and pressure. Our organization was the only queer-focused organization to be invited to testify in front of the National AI Advisory Committee, and the only queer-focused organization and one of the few civil society members of the 280 members of the NIST Artificial Intelligence Safety Institute Consortium. As an organization, we felt pressure to both be excellent advocates for queer people and to represent all queer people and perspectives, analogous to what individuals can feel as the only members of a marginalized community in a group \cite{powell1989women}. This additional pressure contributed to stress and burnout among our organizers. Being the only queer-focused organization also ``pigeon-holed'' policy discussions. Our organization's positions were interpreted as representative of all queer people, rather than productive discussion occurring between multiple queer organizations leading to more nuanced positions representative of more queer people. Our group also struggled with how to relate to others. As the only queer focused organization, we felt pressure to (i) boldly and uncompromisingly represent policy stances that would lead to good futures for queer people; (ii) develop relationships with other organizations; and (iii) make compromises to be invited to the table.

\subsection{Towards effective PD in AI governance}

\paragraph{Coalition building} It is important to increase the power of smaller, volunteer organizations interested in conducting policy work. Coalition building among civil society organizations involved in AISIC allowed for consolidated policy asks. These meetings were two-fold --- serving as a place to openly discuss problems each group faced; and actively connect ideas and viewpoints each group had into shared stances to communicate in the RFCs. The coalition desired to create a ``counter-power'' --- working to shift the power dynamics within AISIC and bring greater validity to socio-technical AI policy approaches. Within the coalition, there was also a focus on transparency to combat frustrations many in the group had communicating with NIST AISI. Multiple civil society groups voiced frustrations of not knowing whether concerns of lacking socio-technical approaches were being considered. Additionally, frustrations came from the siloed nature of working asynchronously, across different working groups and members not being allowed to know AISI guiding discussions. By participating in the coalition we had more ``asking power'' and direct connections to successfully schedule a meeting with NIST AISI leadership to voice these frustrations. A NIST representative noted that they hadn't taken all responses into account when applying revisions and were not aware of concerns that socio-technical considerations had been minimized. Coalition members ultimately chose to shift their efforts towards amplifying why socio-technical approaches should be considered (e.g,. through news op-eds, social media, policy explainers, etc.) as opposed to their original efforts suggesting how. These meetings also inspired our organization’s work to develop a policy explainer. Members from this coalition gave feedback on the explainer and background for how our policy concerns fit within socio-technical AI policy they had already done. 

% Many of discriminatory and representative harms are caused by ``lower tech'' algorithms such as developing an algorithm to determine sexual orientation from Facebook profiles \cite{bhattasali2015machine} and can be addressed by improving data collection and use practices. This broadened focus on ``lower tech'', ``narrow AI'', and generative models allows for the addressal of real-world harms experienced by queer people instead of a sole focus on harms from AI systems.

% Do only the well resourced organizations get to consider the benefits of distributed PD?
\paragraph{Distributed PD in policy design may require prohibitive amounts of volunteer and staff time, especially from smaller organizations.} AI policy has interconnected policy goals affecting data collection, use, and technology. However, this broad focus includes innumerable technologies and policy directions (e.g., AI surveillance modalities, proliferation of censorship and bias in social media algorithms, online hate speech and targeting, discrimination in algorithmically mediated decisions in banking and healthcare). We found this broad set of topics too much for our team to effectively cover, and we struggled to prioritize which topics to advocate on. Ultimately, we decided to focus on general values and principles that would be most applicable to NIST as a non-regulatory agency rather than develop detailed and actionable positions on every issue. Additionally, new policy documents or issues would be raised often and without warning, hindering the ability for detailed and specific proposals to meet the needs of the current moment or opportunity. Some of these issues may be addressed by having multiple and frequent points of distillation \cite{1579648} but the increased person overhead appears to limit this to more resourced organizations. 

\subsection{Applicability In non-US Contexts}

A number of our suggestions to policy makers and organizations are not specific to the US and can be applied in non-US contexts. Notably, \textit{opaque relationships hurt effective AI regulation} and \textit{limited meeting accessibility creates barriers to participation} are not issues unique to the US. Opaque policymaking practices appear in other countries across the world \cite{relly2009perceptions} as does limited meeting accessibility \cite{aguilar2021inclusion}. Similarly, \textit{extractive user-designer relationships} can exist across organizations globally where an organization or government is meeting with a diverse group of participants and/or civil society organizations without meaningfully engaging with these groups or incorporating their expertise.

\section{Limitations and Future Work}
Our team of over $70$ queer folks represented diversity in geography, expertise, and interests, acting as a monitoring mechanism to identify needs affecting queer communities at large. Our team members were primarily based in the US with a few of our members from the Global South and had largely already been involved in AI spaces, thus reflecting associated biases (e.g., comfort, knowledge, and usage of AI). Future work by our organization would aim to elevate the voices of parts of queer communities left out thus far, possibly through community-wide surveys and interviews. At multiple points, we ran into the issue of being an (albeit large) volunteer run, lower resourced organization. We lacked consistently steady manpower to provide multiple channels for feedback, knowledge sharing, and processing suggested in scaled PD literature.  

\section{Conclusion}
In this work, we have presented our organization's efforts to develop and engage in AI policy work while adhering to participatory values as a case study for challenges effective engagement of marginalized communities in AI policy creation. We analyze what specific organizational structures and values were successful, and what led to challenges, both within our organization and the larger policy bodies we participated in. We leverage our experience to provide recommendations to increase participation from marginalized communities in AI policy making, and provide guidance to other organizations working in AI and technology policy. For organizations interested in policy advocacy, mentorship, and coalition building act as important sources to equalize internal power relationships and foster individual agency essential to co-ownership in the co-design process. Additionally, policy makers can foster institutional trust and align AI governance with public will by limiting opaque policy relationships and offering more avenues for direct participation.

\section*{Positionality Statement} 
The authors of this paper have formal training as researchers and practitioners in machine learning, natural language processing, privacy, security, fairness, ethics, and legal research. Many authors also identify as members of marginalized communities and contain expertise in other areas. The authors also have varying levels of experience in policy design, participatory governance, participatory AI design, advocacy, unionizing, and activism. Our shared experiences heavily influence how we write about this work. The focus of this paper is on policy conversations in the US though, the authors are from North America, Asia, and Europe and are knowledgeable about related policy conversations in these locations.

\begin{acks}
    The authors would like to thank Angelina Wang, Arjun Subramonian, B.V. Alaka, Serena Oduro, and others within Queer in AI and Data \& Society for offering valuable feedback on early versions of the paper. We are also indebted to those who offered their time and expertise to guide us through working within the US policy space.
\end{acks}

\section*{Generative AI Usage Statement}
To the best of our knowledge, no authors used generative AI tools during the preparation and editing of the text and figures in this work. Per ACM policy, no generative AI tools were used to generate any text content in part or whole.

% bibliography
\bibliographystyle{ACM-Reference-Format}
\bibliography{references}
\end{document}